\documentclass[preprint,amsmath,amssymb]{revtex4}
\usepackage{graphicx}
\usepackage{dcolumn}
\usepackage{bm}
\usepackage{amsthm}

\newtheorem*{thm}{Theorem}
\renewcommand \baselinestretch{1}

\begin{document}

\title{Collection of Mutually Synchronized Chaotic Systems}

\author{A.I. Lerescu}
\email{a.i.lerescu@rug.nl}
\affiliation{Physics of Nano-Devices, Materials Science Center\\
University of Groningen, The Netherlands }

\author{S. Oancea}
\affiliation{Biophysics Dept., University of Agriculture and
Veterinary Medicine, Iasi 6600, Romania }

\author{I. Grosu}
\email{igrosu@umfiasi.ro} \affiliation{Faculty of Bioengineering,
University of Medicine and Pharmacy, Iasi, 6600, Romania}

\date{\today}

\begin{abstract}
A general explicit coupling for mutual synchronization of two
arbitrary identical continuous systems is proposed. The
synchronization is proved analytically. The coupling is given for
all 19 systems from Sprott's collection. For one of the systems
the numerical results are shown in detail. The method could be
adopted for the teaching of the topic.
\end{abstract}



\maketitle

\section{\label{sec:introduction}Introduction}

Synchronization is a fascinating phenomenon in nature
\cite{strogatz,glass} and a useful one in human
activity~\cite{nijmeijer}. Very simple systems like driven
gravitational pendulums or driven nonlinear LCR electric circuits
can have chaotic behavior. This means that such systems can not be
kept to oscillate in synchrony due to their sensitivity to initial
conditions. To synchronize them an additive coupling have been
tried that is feasible from the engineering point of view: a
constant multiplication of the differences of the states .This
type of driving does not work in general. Highly elaborated and
mathematically based methods are
needed~\cite{grosue,huang,lerescu}. If the above mentioned simple
chaotic systems can be synchronized with a precise coupling that
is not intuitive at all then the coupling of biological cells that
are multivariables nonlinear systems (like millions of neurons
that fire together to control our breathing or the coordinated
firing of thsousands of pacemakers cells in our
hearts~\cite{strogatz}) is hard to be imagined. There are known
many results on several types of synchronization~\cite{kurths}.
Mutual synchronization is implied in the synchronization of
networks~\cite{strogatz,chen}. Here we give, for the first time, a
precise general coupling between two arbitrary identical
oscillators in order to get synchronization. The proposed coupling
is mathematically based and the synchronization is analytically
proved. Numerical results are shown. The paper is organized as
follows. In section~\ref{sec:sych} a coupling is proposed and a
theorem is proved for the synchronization of the two systems.
Section~\ref{sec:numerical} contains detailed calculations for the
Hurwitz matrix and numerical results for one of the systems. The
paper ends with conclusions.

\section{\label{sec:sych}Mutual Synchronization}

Let's consider two identical nonautonomous systems of the form:

\begin{equation}\label{eq1}
\dot{x_{1}}=F(x_{1})+Bu_{1},\qquad y_{1}=Cx_{1},\qquad
x_{1}(0)=x_{10}
\end{equation}

\begin{equation}\label{eq2}
\dot{x_{2}}=F(x_{2})+Bu_{2},\qquad y_{2}=Cx_{2},\qquad
x_{2}(0)=x_{20}
\end{equation}

where $x_{j} \in \Re^{n}$, $u_{j} \in \Re^{n}$, $j=1,2$. For the
prescribed inputs $u_{1}=-\gamma (y_{1}-y_{2})$ and $u_{2}=-\gamma
(y_{2}-y_{1})$, the mutual synchronization of (\ref{eq1}) and
(\ref{eq2}) has been studied in \cite{pogromsky1} using the
powerful concept of passivity \cite{pogromsky2}. Namely, it is
proved that there is a constant $\bar{\gamma}$ such that for
$\gamma > \bar{\gamma}$ the systems (\ref{eq1}) and (\ref{eq2})
synchronize and their dynamics is bounded. Here we consider a
particular case with $B=C=\mathbb{I}$ (identity matrix). The
inputs $u_{1}$ and $u_{2}$ are not prescribed but they result from
the condition of synchronization. In order to avoid the indices
with two figures we adopt the notation: $x_{1}=x$, $x_{2}=y$,
$u_{1}=S(t)u_{x}(x,y)$ and $u_{2}=S(t)u_{y}(x,y)$. In the
following we study the coupled systems:

\begin{equation}\label{eq3}
\dot{x}=F(x)+S(t)u_{x}(x,y),\qquad x(0)=x_{0}
\end{equation}

\begin{equation}\label{eq4}
\dot{y}=F(y)+S(t)u_{y}(x,y),\qquad y(0)=y_{0}
\end{equation}

where $S(t)$ is 0 or 1 as a switch. The above systems will
synchronize if $$\lim_{t \rightarrow \infty} \Arrowvert x(t)-y(t)
\Arrowvert = 0$$ for any $x(0), y(0) \in \mathbb{D} \subset
\Re^{n}$. The proposed couplings are:

\begin{equation}\label{eq5}
u_{x}(x,y)= \left( H-\frac{\mathrm{d}F(s)}{\mathrm{d}s} \right)
 \frac{x-y}{2}
\end{equation}

\begin{equation}\label{eq6}
u_{y}(x,y)= \left( H-\frac{\mathrm{d}F(s)}{\mathrm{d}s} \right)
 \frac{-x+y}{2}
\end{equation}

where $s=\frac{x+y}{2}$ and $H$ is an arbitrary constant Hurwitz
matrix (a matrix with negative real part eigenvalues). With
equations (\ref{eq3}),(\ref{eq4}),(\ref{eq5}) and (\ref{eq6}) we
announce the following Theorem.

\begin{thm}
With $S(t)=1$, the systems (\ref{eq3}) and (\ref{eq4}) with the
coupling (\ref{eq5}) and (\ref{eq6}) will synchronize for any
$x(0)$ and $y(0)$ with $\Arrowvert x(0)-y(0) \Arrowvert$ small
enough.
\end{thm}

\begin{proof}
Along with $s=\frac{x+y}{2}$ we make the notation
$r=\frac{x-y}{2}$. It follows $x=s+r$ and $y=s-r$. Substracting
eq. (\ref{eq4}) from eq. (\ref{eq3}) one has:

\begin{equation}\label{eq7}
\dot{r}=\frac{F(s+r)-F(s-r)}{2}+\left(
H-\frac{\mathrm{d}F(s)}{\mathrm{d}s} \right)r
\end{equation}

We use the Taylor expansions:

\begin{equation}\label{eq8}
F(s+r)=F(s)+\frac{\mathrm{d}F(s)}{\mathrm{d}s}r+
\frac{1}{2}\frac{\mathrm{d}^{2}F(s)}{\mathrm{d}s_{i}\mathrm{d}s_{j}}r_{i}r_{j}+\ldots
\end{equation}

\begin{equation}\label{eq9}
F(s-r)=F(s)+\frac{\mathrm{d}F(s)}{\mathrm{d}s}(-r)+
\frac{1}{2}\frac{\mathrm{d}^{2}F(s)}{\mathrm{d}s_{i}\mathrm{d}s_{j}}r_{i}r_{j}+\ldots
\end{equation}

Using the first three terms in eq. (\ref{eq8}) and eq.
(\ref{eq9}), eq. (\ref{eq7}) becomes:

\begin{equation}\label{eq10}
\dot{r}=Hr
\end{equation}

With $H$ a Hurwitz matrix, eq. (\ref{eq10}) assures that $r(t)
\rightarrow 0$ for any $r(0)$ for which the Taylor expansions eq.
(\ref{eq8}) and eq. (\ref{eq9}) are valid. This means that $x(t)
\rightarrow y(t)$ for any $r(0)=x(0)-y(0)$ with $\Arrowvert
x(0)-y(0) \Arrowvert$ small enough. If $F(x)$ is quadratic then
the Taylor expansions have only three terms and in this case the
eq~(\ref{eq10}) is not anymore an approximation.

\end{proof}

Here we note a significant difference between master-slave
synchronization and mutual synchronization. For master-slave
synchronization \cite{grosue,lerescu}, the error dynamics is
described by an approximate equation like eq~(\ref{eq10}) for any
nonlinear $F(x)$. For mutual synchronization of systems with
$F(x)$ a polynomial up to quadratic terms, eq~(\ref{eq10}) is
exact. This could lead to the conclusion that mutual
synchronization can be achieved in an easier manner. This is not
generally true. A deeper analysis is necessary. In spite of the
fact that the eq~(\ref{eq10}) assures that the two dynamics will
be closer and closer, it is not sure that their dynamics are
bounded. Even if they are bounded they could have large excursions
in the phase space that is very undesirable from the engineering
point of view. Further studies are needed to establish how to
choose the parameter $p$ (or $p_{1}$,$p_{2}$) in the matrix $H$
(see Table I) and the initial conditions in order to have bounded
dynamics of the synchronized systems. Otherwise we can manage this
by switching on/off the coupling like this: $S(t)=1$ when
$\Arrowvert x - y \Arrowvert < \delta$ and $\Arrowvert x
\Arrowvert < R$ where $R$ is the radius of a sphere that contains
the attractor of the system $\dot{x}=F(x)$. In Table II (next
section), we give the numerical values of the parameter and the
initial conditions for which the synchronization was verified
numerically.

Matrix $H$ can be chosen in such a manner that the terms from eq.
(\ref{eq5}) and eq. (\ref{eq6}) to be as simple as possible
\cite{grosue,lerescu} and this depends on the particular form of
$F(x)$. If $\frac{\mathrm{d}F_{i}(s)}{\mathrm{d}s_{k}}$ is
constant then we can choose
$H_{ik}=\frac{\mathrm{d}F_{i}(s)}{\mathrm{d}s_{k}}$ and the
corresponding term $(u_{x})_{ik}$ will be zero. The simplest
coupling in eq. (\ref{eq5}) will be when $F(x)$ contains one
nonlinear term and this contains one variable. In this case the
coupling contains one term (see Table I, below, systems F ,H, I,
J, L, M ,N, P, Q, S). We apply this strategy to all systems from
Sprott's collection ~\cite{sprott}. Table I contains the coupled
eq. (\ref{eq5}) and eq. (\ref{eq6}) for all 19 systems from
Sprott's collection.

\begingroup
\squeezetable
\begin{table}[]\label{tab}
\renewcommand*
\baselinestretch{0.3} \caption{The coupled systems (\ref{eq1}) and
(\ref{eq2}) of the Sprott's collection~\cite{sprott} with
$s_{i}=\frac{x_{i}+y_{i}}{2}$ and $r_{i}=\frac{x_{i}-y_{i}}{2}$
where $i=1,2,3$ and $S(t)=1$.}
\begin{ruledtabular}
\begin{tabular}{|l|l|l|l|}

 System&system x&system y&parameter\\

 \hline

A

&$

\begin{array}{l}

\dot{x}_{1}=x_{2} \\

\dot{x}_{2}=-x_{1}+x_{2}x_{3}+(-1-s_{3})r_{2}+(1-s_{2})r_{3} \\

\dot{x}_{3}=1-x_{2}^{2}+pr_{1}+2s_{2}r_{2} \\

\end{array}$

&$

\begin{array}{l}

\dot{y}_{1}=y_{2} \\

\dot{y}_{2}=-y_{1}+y_{2}y_{3}+(-1-s_{3})(-r_{2})+(1-s_{2})(-r_{3}) \\

\dot{y}_{3}=1-y_{2}^{2}+p(-r_{1})+2s_{2}(-r_{2}) \\

\end{array}$

&$

\begin{array}{l}

 \\

-1<p<0 \\

 \\

\end{array}$\\

 \hline

B

&$

\begin{array}{l}

\dot{x}_{1}=x_{2}x_{3}+(1-s_{3})r_{2}+(p_{1}-s_{2})r_{3} \\

\dot{x}_{2}=x_{1}-x_{2} \\

\dot{x}_{3}=1-x_{1}x_{2}+(p_{2}+s_{2})r_{1}+(1+s_{1})r_{2} \\

\end{array}$

&$

\begin{array}{l}

\dot{y}_{1}=y_{2}y_{3}+(1-s_{3})(-r_{2})+(p_{1}-s_{2})(-r_{3}) \\

\dot{y}_{2}=y_{1}-y_{2} \\

\dot{y}_{3}=1-y_{1}y_{2}+(p_{2}+s_{2})(-r_{1})+(1+s_{1})(-r_{2}) \\

\end{array}$

&$

\begin{array}{l}

p_{1}>1 \\

p_{2}<-1 \\

 \\

\end{array}$\\

\hline

C

&$

\begin{array}{l}

\dot{x}_{1}=x_{2}x_{3}+(p-s_{3})r_{2}-(1+s_{2})r_{3} \\

\dot{x}_{2}=x_{1}-x_{2} \\

\dot{x}_{3}=1-x_{1}^{2}+(1+2s_{1})r_{1} \\

\end{array}$

&$

\begin{array}{l}

\dot{y}_{1}=y_{2}y_{3}+(p-s_{3})(-r_{2})-(1+s_{2})(-r_{3}) \\

\dot{y}_{2}=y_{1}-y_{2} \\

\dot{y}_{3}=1-y_{1}^{2}+(1+2s_{1})(-r_{1}) \\

\end{array}$

&$

\begin{array}{l}

 \\

p<0 \\

 \\

\end{array}$\\

\hline

D

&$

\begin{array}{l}

\dot{x}_{1}=-x_{2} \\

\dot{x}_{2}=x_{1}+x_{3} \\

\dot{x}_{3}=x_{1}x_{3}+3x_{2}^{2}+(p_{2}-s_{3})r_{1}-\\

\qquad -6s_{2}r_{2}+(p_{1}-s_{1})r_{3} \\

\end{array}$

&$

\begin{array}{l}

\dot{y}_{1}=-y_{2} \\

\dot{y}_{2}=y_{1}+y_{3} \\

\dot{y}_{3}=y_{1}y_{3}+3y_{2}^{2}+(p_{2}-s_{3})(-r_{1})-\\

\qquad -6s_{2}(-r_{2})+(p_{1}-s_{1})(-r_{3}) \\

\end{array}$

&$

\begin{array}{l}

p_{1}<0 \\

p_{1}<p_{2} \\

p_{2}<0 \\

\end{array}$\\

\hline

E

&$

\begin{array}{l}

\dot{x}_{1}=x_{2}x_{3}+(p_{1}-s_{3})r_{2}+(p_{2}-s_{2})r_{3} \\

\dot{x}_{2}=x_{1}^{2}-x_{2}+(1-2s_{1})r_{1} \\

\dot{x}_{3}=1-4x_{1}x_{2}+(-4+4s_{2})r_{1}+4s_{1}r_{2} \\

\end{array}$

&$

\begin{array}{l}

\dot{y}_{1}=y_{2}y_{3}+(p_{1}-s_{3})(-r_{2})+(p_{2}-s_{2})(-r_{3}) \\

\dot{y}_{2}=y_{1}^{2}-y_{2}+(1-2s_{1})(-r_{1}) \\

\dot{y}_{3}=1-4y_{1}y_{2}+(-4+4s_{2})(-r_{1})+4s_{1}(-r_{2}) \\

\end{array}$

&$

\begin{array}{l}

p_{1}<0 \\

p_{2}>0 \\

 \\

\end{array}$\\

\hline

F

&$

\begin{array}{l}

\dot{x}_{1}=x_{2}+x_{3} \\

\dot{x}_{2}=-x_{1}+0.5x_{2} \\

\dot{x}_{3}=x_{1}^{2}-x_{3}+(p-2s_{1})r_{1} \\

\end{array}$

&$

\begin{array}{l}

\dot{y}_{1}=y_{2}+y_{3} \\

\dot{y}_{2}=-y_{1}+0.5y_{2} \\

\dot{y}_{3}=y_{1}^{2}-y_{3}+(p-2s_{1})(-r_{1}) \\

\end{array}$

&$

\begin{array}{l}

 \\

p \in (-2,-0.75) \\

 \\

\end{array}$\\

\hline

G

&$

\begin{array}{l}

\dot{x}_{1}=0.4x_{1}+x_{3} \\

\dot{x}_{2}=x_{1}x_{3}-x_{2}-s_{3}r_{1}+(p-s_{1})r_{3} \\

\dot{x}_{3}=-x_{1}+x_{2} \\

\end{array}$

&$

\begin{array}{l}

\dot{y}_{1}=0.4y_{1}+y_{3} \\

\dot{y}_{2}=y_{1}y_{3}-y_{2}-s_{3}(-r_{1})+(p-s_{1})(-r_{3}) \\

\dot{y}_{3}=-y_{1}+y_{2} \\

\end{array}$

&$

\begin{array}{l}

 \\

p \in (-2.5,-0.6) \\

 \\

\end{array}$\\

\hline

H

&$

\begin{array}{l}

\dot{x}_{1}=-x_{2}+x_{3}^{2}+(p-2s_{3})r_{3} \\

\dot{x}_{2}=x_{1}+0.5x_{2} \\

\dot{x}_{3}=x_{1}-x_{3} \\

\end{array}$

&$

\begin{array}{l}

\dot{y}_{1}=-y_{2}+y_{3}^{2}+(p-2s_{3})(-r_{3}) \\

\dot{y}_{2}=y_{1}+0.5y_{2} \\

\dot{y}_{3}=y_{1}-y_{3} \\

\end{array}$

&$

\begin{array}{l}

 \\

p \in (-2,-0.75) \\

 \\

\end{array}$\\

\hline

I

&$

\begin{array}{l}

\dot{x}_{1}=-0.2x_{2} \\

\dot{x}_{2}=x_{1}+x_{3} \\

\dot{x}_{3}=x_{1}+x_{2}^{2}-x_{3}+(p-2s_{2})r_{2} \\

\end{array}$

&$

\begin{array}{l}

\dot{y}_{1}=-0.2y_{2} \\

\dot{y}_{2}=x_{1}+y_{3} \\

\dot{y}_{3}=x_{1}+y_{2}^{2}-y_{3}+(p-2s_{2})(-r_{2}) \\

\end{array}$

&$

\begin{array}{l}

 \\

p<-0.2 \\

 \\

\end{array}$\\

\hline

J

&$

\begin{array}{l}

\dot{x}_{1}=2x_{3} \\

\dot{x}_{2}=-2x_{2}+x_{3} \\

\dot{x}_{3}=-x_{1}+x_{2}+x_{2}^{2}+(p-1-2s_{2})r_{2} \\

\end{array}$

&$

\begin{array}{l}

\dot{y}_{1}=2y_{3} \\

\dot{y}_{2}=-2y_{2}+y_{3} \\

\dot{y}_{3}=-y_{1}+y_{2}+y_{2}^{2}+(p-1-2s_{2})(-r_{2}) \\

\end{array}$

&$

\begin{array}{l}

 \\

p<0 \\

 \\

\end{array}$\\

\hline

K

&$

\begin{array}{l}

\dot{x}_{1}=x_{1}x_{2}-x_{3}-s_{2}r_{1}+(p-s_{1})r_{2} \\

\dot{x}_{2}=x_{1}-x_{2} \\

\dot{x}_{3}=x_{1}+0.3x_{3} \\

\end{array}$

&$

\begin{array}{l}

\dot{y}_{1}=y_{1}y_{2}-y_{3}-s_{2}(-r_{1})+(p-s_{1})(-r_{2}) \\

\dot{y}_{2}=y_{1}-y_{2} \\

\dot{y}_{3}=y_{1}+0.3y_{3} \\

\end{array}$

&$

\begin{array}{l}

 \\

p \in (-3.3,-0.51) \\

 \\

\end{array}$\\

\hline

L

&$

\begin{array}{l}

\dot{x}_{1}=x_{2}+3.9x_{3} \\

\dot{x}_{2}=0.9x_{1}^{2}-x_{2}+(p-1.8s_{1})r_{1} \\

\dot{x}_{3}=1-x_{1} \\

\end{array}$

&$

\begin{array}{l}

\dot{y}_{1}=y_{2}+3.9y_{3} \\

\dot{y}_{2}=0.9y_{1}^{2}-y_{2}+(p-1.8s_{1})(-r_{1}) \\

\dot{y}_{3}=1-y_{1} \\

\end{array}$

&$

\begin{array}{l}

 \\

p<0 \\

 \\

\end{array}$\\

\hline

M

&$

\begin{array}{l}

\dot{x}_{1}=-x_{3} \\

\dot{x}_{2}=-x_{1}^{2}-x_{2}+(p+2s_{1})r_{1} \\

\dot{x}_{3}=1.7+1.7x_{1}+x_{2} \\

\end{array}$

&$

\begin{array}{l}

\dot{y}_{1}=-y_{3} \\

\dot{y}_{2}=-y_{1}^{2}-y_{2}+(p+2s_{1})(-r_{1}) \\

\dot{y}_{3}=1.7+1.7y_{1}+y_{2} \\

\end{array}$

&$

\begin{array}{l}

 \\

p \in (-1.7,0) \\

 \\

\end{array}$\\

\hline

N

&$

\begin{array}{l}

\dot{x}_{1}=-2x_{2} \\

\dot{x}_{2}=x_{1}+x_{3}^{2}+(p-2s_{3})r_{3} \\

\dot{x}_{3}=1+x_{2}-2x_{3} \\

\end{array}$

&$

\begin{array}{l}

\dot{y}_{1}=-2y_{2} \\

\dot{y}_{2}=y_{1}+y_{3}^{2}+(p-2s_{3})(-r_{3}) \\

\dot{y}_{3}=1+y_{2}-2y_{3} \\

\end{array}$

&$

\begin{array}{l}

 \\

p < 0 \\

 \\

\end{array}$\\

\hline

O

&$

\begin{array}{l}

\dot{x}_{1}=x_{2} \\

\dot{x}_{2}=x_{1}-x_{3} \\

\dot{x}_{3}=x_{1}+x_{1}x_{3}+2.7x_{2}-s_{3}r_{1}+(p-s_{1})r_{3} \\

\end{array}$

&$

\begin{array}{l}

\dot{y}_{1}=y_{2} \\

\dot{y}_{2}=y_{1}-y_{3} \\

\dot{y}_{3}=y_{1}+y_{1}y_{3}+2.7y_{2}-s_{3}(-r_{1})+(p-s_{1})(-r_{3}) \\

\end{array}$

&$

\begin{array}{l}

 \\

p \in (-1,-0.37) \\

 \\

\end{array}$\\

\hline

P

&$

\begin{array}{l}

\dot{x}_{1}=2.7x_{2}+x_{3} \\

\dot{x}_{2}=-x_{1}+x_{2}^{2}+(p-2s_{2})r_{2} \\

\dot{x}_{3}=x_{1}+x_{2} \\

\end{array}$

&$

\begin{array}{l}

\dot{y}_{1}=2.7y_{2}+y_{3} \\

\dot{y}_{2}=-y_{1}+y_{2}^{2}+(p-2s_{2})(-r_{2}) \\

\dot{y}_{3}=y_{1}+y_{2} \\

\end{array}$

&$

\begin{array}{l}

 \\

p \in (-1,\frac{10}{27}) \\

 \\

\end{array}$\\

\hline

Q

&$

\begin{array}{l}

\dot{x}_{1}=-x_{3} \\

\dot{x}_{2}=x_{1}-x_{2} \\

\dot{x}_{3}=3.1x_{1}+x_{2}^{2}+0.5x_{3}+(p-2s_{2})r_{2} \\

\end{array}$

&$

\begin{array}{l}

\dot{y}_{1}=-y_{3} \\

\dot{y}_{2}=y_{1}-y_{2} \\

\dot{y}_{3}=3.1y_{1}+y_{2}^{2}+0.5y_{3}+(p-2s_{2})(-r_{2}) \\

\end{array}$

&$

\begin{array}{l}

 \\

p \in (-3.1,-1.8) \\

 \\

\end{array}$\\

\hline

R

&$

\begin{array}{l}

\dot{x}_{1}=0.9-x_{2} \\

\dot{x}_{2}=0.4+x_{3} \\

\dot{x}_{3}=x_{1}x_{2}-x_{3}+(1-s_{2})r_{1}+(p-s_{1})r_{2} \\

\end{array}$

&$

\begin{array}{l}

\dot{y}_{1}=0.9-y_{2} \\

\dot{y}_{2}=0.4+y_{3} \\

\dot{y}_{3}=y_{1}y_{2}-y_{3}+(1-s_{2})(-r_{1})+(p-s_{1})(-r_{2}) \\

\end{array}$

&$

\begin{array}{l}

 \\

p < -1 \\

 \\

\end{array}$\\

\hline

S

&$

\begin{array}{l}

\dot{x}_{1}=-x_{1}-4x_{2} \\

\dot{x}_{2}=x_{1}+x_{3}^{2}+(p-2s_{3})r_{3} \\

\dot{x}_{3}=1+x_{1} \\

\end{array}$

&$

\begin{array}{l}

\dot{y}_{1}=-y_{1}-4y_{2} \\

\dot{y}_{2}=y_{1}+y_{3}^{2}+(p-2s_{3})(-r_{3}) \\

\dot{y}_{3}=1+y_{1} \\

\end{array}$

&$

\begin{array}{l}

 \\

p \in (0,1) \\

 \\

\end{array}$\\

\end{tabular}
\end{ruledtabular}
\end{table}
\endgroup

\section{\label{sec:numerical}Numerical results}

We give detailed calculations for the system S from the Sprott's
collection (see Table I). The system in $x$ is:

\begin{equation}\label{eq11}
\begin{array}{lcl}
\dot{x}_{1} & = & -x_{1} + 4x_{2} \\
\dot{x}_{2} & = & x_{1} + x_{3}^{2} \\
\dot{x}_{3} & = & 1 + x_{1}
\end{array}
\end{equation}

The Jacobian is:

\begin{equation}\label{eq12}
\frac{\mathrm{d}F(s)}{\mathrm{d}s} = \left (
\begin{array}{ccc}
-1 & -4 & 0 \\
1 & 0 & 2s_{3} \\
1 & 0 & 0\\
\end{array}
\right )
\end{equation}

We can choose matrix $H$ as:

\begin{equation}\label{eq13}
H = \left (
\begin{array}{ccc}
-1 & -4 & 0 \\
1 & 0 & p \\
1 & 0 & 0\\
\end{array}
\right )
\end{equation}

The characteristic equation is:

\begin{equation}\label{eq14}
\lambda^{3} + a_{1} \lambda^{2} + a_{2} \lambda + a_{3} = 0
\end{equation}

where

\begin{equation}\label{eq15}
a_{1}=1, \quad a_{2}=4, \quad a_{3}=4p.
\end{equation}

The Ruth-Hurwitz conditions (for eq. (\ref{eq14})) are:

\begin{equation}\label{eq16}
a_{1} > 0, a_{1}a_{2} - a_{3} > 0, a_{3} > 0.
\end{equation}

\begin{figure}[t]\label{figur}
\includegraphics[width=10cm]{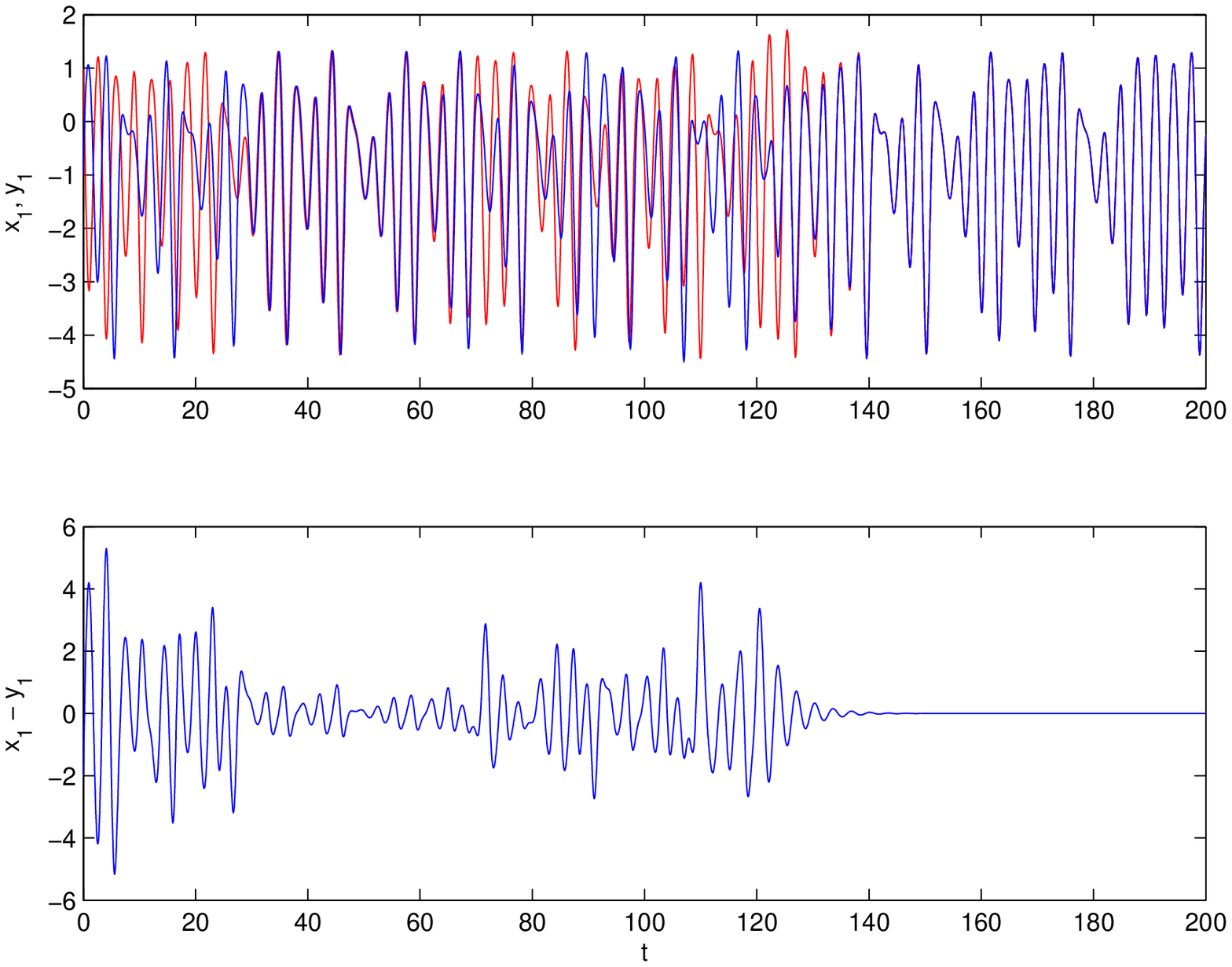}
\caption{Numerical results. $S(t)=0$ for $t<120$ and $S(t)=1$ for
$t>120$ for the coupled systems (\ref{eq18}) and (\ref{eq19}).
Initial conditions are $x_{1}(0)=x_{2}(0)=x_{3}(0)=-1$ and
$y_{1}(0)=y_{2}(0)=y_{3}(0)=1$. Top figure represents $x_{1}$ and
$y_{1}$ and the bottom figure is the error function $x_{1}-y_{1}$.
It takes about 15 time units since the mutual coupling is switched
on (at $t=120$) until the synchronization is achieved.}
\end{figure}

This gives for the parameter $p$ the condition $0 < p < 1$. In
this case we have:

\begin{eqnarray}\label{eq17}
\begin{array}{ll}
u_{x}^{1} = 0 & u_{y}^{1} = 0 \\
u_{x}^{2} = (p-2s_{3})r_{3} & u_{y}^{2} = (p-2s_{3})(-r_{3}) \\
u_{x}^{3} = 0 & u_{y}^{3} = 0 \\
\end{array}
\end{eqnarray}

The two coupled systems follow now as (see Figure 1 for numerical
simulation):

\begin{eqnarray}\label{eq18}
\nonumber
  \dot{x_{1}} &=& -x_{1}+4x_{2} \\
  \dot{x_{2}} &=& x_{1}+4x_{3}^{2}+S(t)(p-2s_{3})r_{3} \\
\nonumber
  \dot{x_{3}} &=& 1+x_{1} \\
  \nonumber
\end{eqnarray}

\begin{eqnarray}\label{eq19}
\nonumber
  \dot{y_{1}} &=& -y_{1}+4y_{2} \\
  \dot{y_{2}} &=& y_{1}+4y_{3}^{2}+S(t)(p-2s_{3})(-r_{3}) \\
\nonumber
  \dot{y_{3}} &=& 1+y_{1} \\
  \nonumber
\end{eqnarray}

where $s_{3}=\frac{x_{3}+y_{3}}{2}$ and
$r_{3}=\frac{x_{3}-y_{3}}{2}$.

In the same manner can be written the couplings for all 19 systems
from Sprott's collection. If the conditions (\ref{eq16}) can not
be fulfilled with one parameter $p$ then a second parameter should
be introduced (see Table I, systems B, C, D, E, O, R). The Hurwitz
matrix should be chosen the same as in~\cite{lerescu}. The choice
of $H$ is not unique for systems that have several terms in the
coupling term. And, as was mentioned in the previous section, the
theorem does not assures that the dynamics of the coupled systems
is bounded.

In Table II we present numerical values for the parameter $p$ or
$p_{1},p_{2}$ (see Table I) and initial conditions for which the
synchronization was verified numerically. It can be observed that
it was not difficult to find such numerical values. They are
rather homogeneous. When the dynamics was far from the dynamics of
the original system $\dot{x}=F(x)$ for $x(0)=-y(0)=(1,1,1)$ then
we changed to the initial conditions $x(0)=-y(0)=(0.1,0.1,0.1)$.
Further studies are needed to find analytic conditions for the
matrix Hurwitz H and the initial conditions that assure a bounded
dynamics for the coupled system.

\begingroup
\squeezetable
\begin{table}[]\label{tabii}
\renewcommand*
\baselinestretch{0.3} \caption{Numerical values of the parameter
$p$ or $p_{1},p_{2}$ and initial conditions for the systems in
Table I for which the synchronization was verified numerically.}
\begin{ruledtabular}
\begin{tabular}{|c|c|c|c|c|c|c|c|c|c|c|c|c|c|c|c|c|c|c|c|}

 System&A&B&C&D&E&F&G&H&I&J&K&L&M&N&O&P&Q&R&S\\

 \hline

$
\begin{array}{c}

x_{i}(0)=-y_{i}(0) \\

i=1,2,3 \\

\end{array}$

&1&1&1&0.1&1&0.1&0.1&1&1&1&1&1&0.1&1&0.1&0.1&1&1&1\\

 \hline

p&-0.1&&-1&&&-1&-1&-1&-10&-1&-1&-1&-1&-1&-0.5&-0.5&-2&-2&0.5\\

 \hline

$
\begin{array}{c}

p_{1} \\

p_{2} \\

\end{array}$

&&

$
\begin{array}{c}

2 \\

-2 \\

\end{array}$

&&

$
\begin{array}{c}

-0.1 \\

-0.05 \\

\end{array}$

&

$
\begin{array}{c}

-1 \\

1 \\

\end{array}$

&&&&&&&&&&&&&&\\

\end{tabular}
\end{ruledtabular}
\end{table}
\endgroup

Using the same notation like the Table I and the Hurwitz matrix
found in \cite{lerescu}, two Lorenz systems mutually synchronized
look like:

\begin{eqnarray}\label{eq20}
\nonumber
\dot{x}_{1} & = & s(-x_{1} + x_{2}) \\
\dot{x}_{2} & = & rx_{1} - x_{2}-x_{1}x_{3}+S(t)(s_{1}x_{3}+(p+s_{3})r_{1}) \\
\nonumber \dot{x}_{3} & = & -bx_{3} + x_{1}x_{2} - S(t)(s_{1}r_{2}
+ s_{2}r_{1})
\nonumber
\end{eqnarray}
\begin{eqnarray}\label{eq21}
\nonumber
\dot{y}_{1} & = & s(-y_{1} + y_{2}) \\
\dot{y}_{2} & = & ry_{1} - y_{2}-y_{1}y_{3} + S(t)(-s_{1}y_{3}+(p+s_{3})(-r_{1})) \\
\nonumber \dot{y}_{3} & = & -by_{3} + y_{1}y_{2} + S(t)(s_{1}r_{2}
+ s_{2}r_{1})
\nonumber
\end{eqnarray}

For (s,r,b,p) = (16, 45.6, 4, -60) and $x(0)=-y(0)=(1,1,1)$, the
synchronization was verified numerically. In \cite{pogromsky1},
the mutual synchronization of two Lorenz systems was obtained by
using one term in the first equations and any intial conditions.

Also for the simplest chaotic sytems $\dddot{x} + a \ddot{x} -
\dot{x}^{2} + x = 0$, the coupled systems that will synchronize
look like: $\dddot{x} + a \ddot{x} - \dot{x}^{2} + x +
(p+(\dot{x}+\dot{y}))(\dot{x}-\dot{y})= 0$ and $\dddot{y} + a
\ddot{y} - \dot{y}^{2} + y +
(p+(\dot{x}+\dot{y}))(-\dot{x}+\dot{y})= 0$ and for $(a,p) =
(2,-1)$ and
$(x(0),\dot{x}(0),\ddot{x}(0))=-(y(0),\dot{y}(0),\ddot{y}(0))=(1,1,1)$
the synchronization was checked numerically.

\section{\label{sec:conclusions}Conclusions}

A precise analytical scheme for mutual synchronization is
proposed. The method is mathematical rigorous in terms of choosing
the expression of coupling and indications are given how to avoid
unbounded dynamics of the coupled systems.

\begin{acknowledgments}
The authors would like to thank the referees for their valuable
comments and suggestions.
\end{acknowledgments}

\bibliography{tabletryr}

\end{document}